\documentclass[preprint,prl]{revtex4}
\usepackage{graphicx,verbatim}
\usepackage{amsmath}
\usepackage{amssymb}
\usepackage{amsfonts}
\usepackage{dsfont}

\newcommand{\ket}[1]{| {#1} \rangle }

\begin{document}

\title{Observation of cooperatively enhanced atomic dipole forces from NV centers in optically trapped nanodiamonds}
\author{Mathieu L. Juan$^{1,2,\ast}$, Carlo Bradac$^{1,2,\ast}$, Benjamin Besga$^{1,2}$, Gavin Brennen$^{1,2}$, Gabriel Molina-Terriza$^{1,2}$ and Thomas Volz$^{1,2}$}

\affiliation{$^1$ Department of Physics \& Astronomy, Macquarie
University, NSW 2109, Australia
\\$^2$ ARC Centre of
Excellence for Engineered Quantum Systems, \\Macquarie University,
NSW 2109, Australia
\\ \textnormal{$^\ast$ These authors contributed equally to this work.}}
\maketitle


{\bf Since the early work by Ashkin in 1970 \cite{Ashkin86}, optical
trapping has become one of the most powerful tools for manipulating
small particles, such as micron sized beads~\cite{Neuman04} or single atoms~\cite{Schlosser2001}. The optical trapping mechanism is based on
the interaction energy of a dipole and the electric field of
the laser light. In atom trapping, the dominant contribution typically comes from the allowed optical transition closest
to the laser wavelength, whereas for mesoscopic particles it is given by the bulk polarizability of the
material. These two different
regimes of optical trapping have coexisted for decades without any
direct link, resulting in two very different contexts of
applications: one being the trapping of small objects mainly in
biological settings \cite{Grier03}, the other one being dipole traps
for individual neutral atoms \cite{Grimm2000} in the field of quantum optics. Here we show that for
nanoscale diamond crystals containing artificial atoms, so-called
nitrogen vacancy (NV) color centers, both regimes of optical
trapping can be observed at the same time even in a noisy liquid
environment. For
wavelengths in the vicinity of the zero-phonon line transition of the color
centers, we observe a significant modification ($10\%$) of the
overall trapping strength. Most remarkably, our experimental findings suggest that owing to the large number of artificial atoms, collective effects greatly contribute to the observed trapping strength modification. Our approach adds the powerful atomic-physics
toolbox to the field of nano-manipulation.}

Whenever a polarizable particle is exposed to light, the
electromagnetic field induces an optical dipole moment which in turn
leads to an interaction energy that scales with the field intensity.
This interaction energy results in optical forces that ultimately allow for
spatial manipulation of the particle. Using a semi-classical
approach the optical forces can be derived as
\cite{Grynberg2010}:
\begin{equation}
\label{eq:force} \mathbf{F}(\mathbf{r}) = \epsilon_0 \alpha '
\frac{E_0(\mathbf{r})}{2} \nabla{E_0(\mathbf{r})} - \epsilon_0
\alpha '' \frac{E_0^2(\mathbf{r})}{2} \nabla{\phi(\mathbf{r})}
\end{equation}
where the polarizability $\alpha$ of the particle has been split
into real and imaginary parts $\alpha = \alpha ' + i \alpha ''$, $
\epsilon_0$ is the dielectric permittivity of vacuum, and the
incident time-averaged field amplitude and phase are $E_0$ and $\phi$,
respectively. The real part of the polarizability gives rise to the
so-called dipole force associated with a conservative trapping
potential. In contrast, the imaginary part of $\alpha$ leads to
dissipative resonant scattering forces proportional to the gradient
of the phase of the field. Due to its dissipative nature, the
resonant scattering force plays a key role in atom cooling
\cite{Grynberg2010}. In the context of this work, we can neglect the
resonant scattering term by utilizing a Gaussian standing wave (GSW) trap~\cite{Zemanek1999}.

For an isotropic and homogeneous object, the polarizability $\alpha$
is directly related to the refractive index of the material and is
typically a slowly varying function of wavelength. Conversely, in the case of a two-level quantum system (e.g. a single atom in free space) the dipole force exhibits a strong dependence on the
trapping wavelength through the resonant nature of the
polarizability. In particular, it changes sign at the transition frequency, $\omega_0$, and
becomes repulsive for blue detunings ($\omega > \omega_0$).
Experiments on ultracold neutral atoms routinely exploit this strong
detuning dependence for creating complex potential landscapes
\cite{Grimm2000}. In the following, we report on the observation of near-resonant optical forces from an ensemble of artificial atoms
embedded in a nanocrystal.

\begin{figure}
    \includegraphics[scale=1]{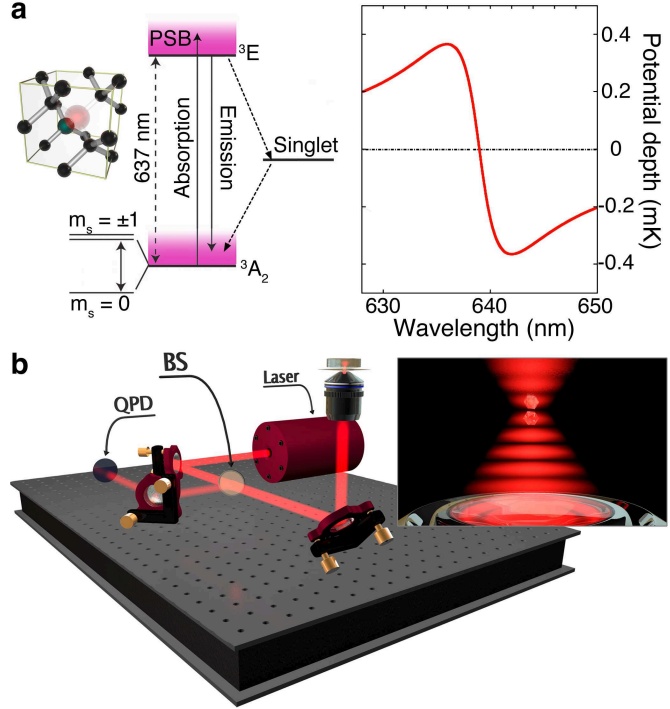}
    \centering
\caption{\textbf{Trapping nanodiamonds.} \textbf{a,}
Highly-irradiated nanodiamonds contain many nitrogen vacancy centers
which exhibit a strong optical transition at around 637~nm (left
panel). The level scheme displays the relevant optical and non-radiative transitions (for spectroscopic notation see e.g.~\cite{Doherty2013}). A single NV center trapped in the center of a focussed
Gaussian beam (numerical aperture 1.2, P=~4~mW) experiences
a dispersive atom-like trapping potential due to the
resonant dipole force (right panel). \textbf{b,} Experimental setup
with the trapping laser focussed through a high numerical aperture
objective into a micro-fluidic chamber producing a strong optical
trap. The top of the micro-fluidic chamber consists of a mirror
which results in a standing wave, improving the trapping efficiency
and minimizing the scattering force. The light backscattered from the system is
collected via a beam splitter (BS) and sent to a quadrant
photo-detector (QPD) to track the position of the particle.} \label{fig1}
\end{figure}

Nanodiamonds (NDs) containing color centers are excellent candidate
nanoparticles for observing atom-like trapping resonances. The color
centers act like artificial atoms exhibiting sharp optical transitions. In particular, the nitrogen-vacancy center
(NV), consisting of a nitrogen atom and an adjacent vacancy site
(see Figure \ref{fig1}a left), has attracted a lot of interest
over the past decade. In its most stable form, the negatively
charged NV$^{-}$, it exhibits outstanding spin-optical properties
which persist even up to room-temperature \cite{Doherty2013}.
Consequently, the NV$^{-}$ has proven to be highly suitable as a
solid-state spin qubit \cite{Jelezko2004b} and nanoscale magnetic
sensor \cite{Balasubramanian2008,Maze2008}. Here we are mainly
interested in its optical properties. The NV$^-$ displays stable
single-photon emission \cite{Brouri2000,Kurtsiefer2000} with a sharp
zero-phonon line (ZPL) in bulk at 637 nm followed by well-defined
vibronic side bands \cite{Davies1976}. At room temperature, most photons are emitted into these sidebands and only a fraction into the ZPL (typically $4\%$~\cite{Faraon2011}). Note that, mainly due to
strain, the ZPL position in NDs can shift considerably. In our
experiment, we measured the average ZPL to be 639~nm. Due to their
strong and stable fluorescence, NV$^-$ centres hosted in
nanodiamonds have been used as biolabels for high-resolution,
real-time and low-disruption imaging of living cells
\cite{McGuinness2011} and as carriers for drugs and biomolecules
\cite{Alhaddad2011}. Previous investigations on liquid trapping
\cite{Horowitz2012,Geiselmann2013} and levitating NDs
\cite{Neukirch2013} used laser light at 1064~nm which is far away
from the ZPL at 637~nm. None of these experiments reported any
effects {due to the presence of the NV$^-$ centers on the external degrees of freedom of the nanodiamond in the
optical trap}.

In our experiment, we create a GSW trap near 639~nm by focussing a Gaussian laser beam on
a silver-coated mirror. {The GSW} provides a stronger trap along the direction of the standing wave compared a to conventional focused Gaussian beam, and allows one to neglect scattering forces. The mirror forms the top of a
static micro-fluidic chamber that contains the NDs suspended in
deionized water (see Figure \ref{fig1}b). The laser sources are a set of temperature-stabilized laser diodes operating
at different wavelengths detuned with respect to 639~nm (see
Materials and Methods). In addition, a pulsed green laser (532~nm)
serves as a weak re-pump to counteract resonant ionization to the
neutral NV$^{0}$ state \cite{Aslam2013}. The trap depth itself is
hard to measure directly in optical tweezers in liquid. Instead, by
measuring the corner frequency \cite{Neuman04} (see Materials and Methods), we obtain the trap
stiffness, $\kappa$, which corresponds to the second derivative of
the trapping potential. The nanodiamonds
are in the strongly overdamped regime and their displacement from
the trap center is small compared to the beam waist. As a result,
the particles mainly probe the harmonic part of the potential near
the trap minimum $U \sim U(0)+ \kappa x^2$ \cite{Gieseler2013}. In
order to obtain the stiffness experimentally, the position of
the trapped particle is recorded using a quadrant photodiode. The stiffness can then be obtained from a Lorentzian fit of the power spectrum density of the signal~\cite{Neuman04} (see Materials and Methods). Due to its dependence on the exact size of
the trapped ND, $\kappa$ is not a good observable when comparing the
results obtained for different NDs. In order to circumvent this
problem, we extract a relative value for $\kappa$ normalized to the
measured trapping stiffness at a reference wavelength
($\lambda_{ref} = 639.13$~nm) for each ND separately (see
Materials and Methods).

\begin{figure}
    \includegraphics[scale=1]{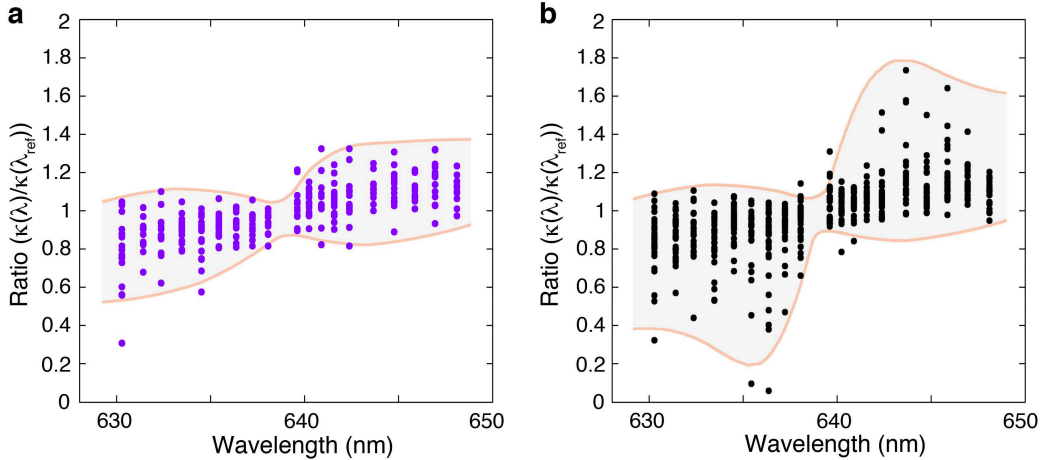}
    \centering
\caption{\textbf{Measured relative trap stiffness.} {\textbf{a,}
Relative trap stiffness ($\kappa(\lambda)/\kappa(\lambda_{ref})$) for NDs
containing only few NVs. The symmetric spread in the measured trap
stiffness for a give wavelength is due to experimental noise. \textbf{b,} Relative trap stiffness for NDs containing a large number of NV
centers. The data scattering is strongly asymmetric with $\kappa$-ratios much lower than 1 for wavelengths below $\lambda_{ref}$ and higher than 1 for wavelengths above $\lambda_{ref}$. \textbf{a, b} The shaded areas are guides to the eye and indicate the
range of data scattering.}} \label{fig2}
\end{figure}

In order to obtain a significant resonant trapping effect, we use NDs with a
high concentration of NV$^-$ centers~\cite{Fu2007}. We first
characterized the NDs using a home-built combined confocal/AFM
microscope setup \cite{Bradac2009}. The investigated NV$^-$ centers
show a ZPL centered at {$(639.08\pm 0.65)$~nm} with an average spectral width of
{$(2.09\pm 0.55)$~nm} (see Extended Data Fig. 1).
The ZPL width of individual nanodiamonds is a convolution of a Gaussian and a Lorentzian distributions (Voigt profile). Assuming a dephasing rate at room temperature of approximately $2\pi\times1$~THz~\cite{Fu2009}, we extracted the width of the underlying Gaussian distribution due to the variation of NV ZPL frequency within a ND to be {$\sigma_{ZPL} = (1.82\pm 0.55)\,\mathrm{nm}$}.
We also measured the size distribution of the nanocrystals in liquid using dynamic light
scattering and found an average size of {$(150 \pm 23)$~nm}. The expected
average number of NVs per ND is $\left< NV \right>
\approx 9,500$~\cite{Fu2007}. {A sample of NDs with a similar size, $(168 \pm 31)$~nm, but with a much lower concentration of NV centers was used as a reference.}

To characterize the contribution of the NVs to the trap stiffness, we measured the corner frequency on a number of
different NDs for a given set of wavelengths {for the two ND samples (low and high NV center concentration)}. We then extracted the
trap stiffness for each ND separately and applied a statistical
analysis in order to {systematically discard} unwanted events such as the trapping of multiple NDs or standing wave hopping (see Supplementary Material). The resulting trap stiffness ratios as a
function of laser wavelength for the reference NDs are displayed in Figure~\ref{fig2}a. This measurement serves as a reference as we do not expect the resonant trapping
forces to have a measurable effect for this sample. The monotonic trend is attributed to chromatic aberrations which are aggravated by the standing wave trap. In contrast, Figure~\ref{fig2}b displays the results for
the ND sample with high NV$^-$ density. As in the case of the reference sample, the underlying monotonic trend is clearly visible. However, the high density data clearly shows a strongly asymmetrical distribution
with $\kappa$-ratio values much lower than 1 for wavelengths below
$\lambda_{ref}$ and values larger than 1 for wavelengths larger than
$\lambda_{ref}$. The extreme values are attributed to NDs with larger number of NV centers leading to a significant contribution of the NVs to the trap stiffness.

\begin{figure}
    \includegraphics[scale=1]{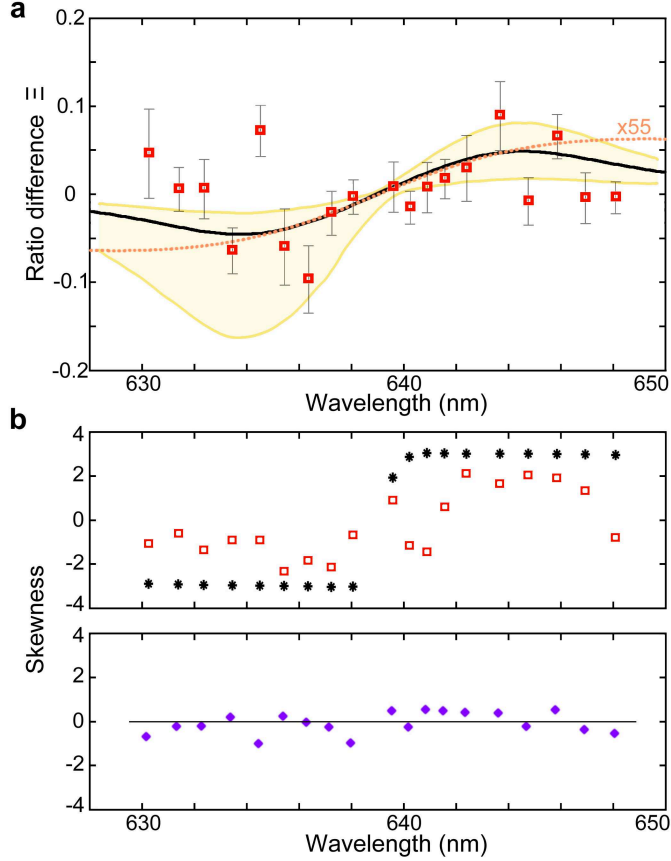}
\centering
\caption{\textbf{Cooperative dipole force.} \textbf{a,} Difference of the mean values from figures
(\ref{fig2}\textbf{a}) and (\ref{fig2}\textbf{b}). This difference, $\Xi = \left. \kappa(\lambda)/\kappa(\lambda_{ref}) \right|_{highNV}- \left. \kappa(\lambda)/\kappa(\lambda_{ref})\right|_{lowNV}$, is a close approximation of the ratio of the stiffness from the NV centres, $\kappa_{NVs}(\lambda)$,  over the stiffness from the diamond matrix, $\kappa_{Diamond}(\lambda)$ (see Supplementary Information). The standard error is indicated in grey. The black curve shows the theoretically predicted stiffness ratio accounting for collective effects on the force obtained for collective sub-ensembles with a spectral width of $100\,\textnormal{GHz}$. {For comparison, the dashed line is obtained by considering independent NVs (magnified 55 times). In addition, our Monte Carlo calculation predicts that in the absence of noise, $90\%$ of the experimental values for the difference ratio $\Xi$ should lie within the shaded area.} \textbf{b,} Skewness as a function of wavelength. (top panel) Experimental skewness (open squares) from high density NV sample measurements along with the simulation results from our Monte-Carlo including collective effects (solid stars). (bottom panel) experimental skewness from the low density NV sample {(solid diamonds)}. Here the skewness displays no obvious deviation from zero.} \label{fig3}
\end{figure}

Next, we extract the mean value from both data sets and
plot the difference of the mean values as a function of wavelength
(Figure \ref{fig3}a). Due to the choice of reference wavelength, this difference of mean values gives access to the ratio of the stiffness arising from the NV ensemble to the stiffness from the bare ND matrix (see Supplementary Material). The dispersive trend is clearly visible, corresponding to a stiffness ratio lower than 1 at wavelengths below the ZPL and larger than 1 for wavelengths above, with a magnitude of up to 10$\%$.
{Most remarkably, these experimental results cannot be accounted for only by assuming independent NVs: The dashed line in Figure~\ref{fig3}a displays the expected trap stiffness from a calculation that assumes independent NV centers and takes the average values for the ZPL width, ZPL position and number of NV centers per ND. Note that the curve has been magnified by 55 times for plotting. Clearly, assuming the NV centers to act independently does not reproduce our experimental findings. Besides the magnitude being off by almost two orders of magnitude, the shape of the curve also does not match the observations. {In order} to fully capture the experimental results, {we therefore} have to consider collective effects~\cite{Dicke1958} between the NV centers within a single nanocrystal.}

Cooperative effects between NV centers have a significant impact on the dipole forces by increasing the spontaneous decay rate and modifying the steady state population. {The presence of collective effects in our high NV density NDs was independently confirmed through a set of scattering experiments where sub-nanosecond cooperative decay rates were observed~\cite{Bradac2015}.} Due to the {large variation in ZPL position in a single ND}, only sub-domains of NVs within narrow frequency windows are expected to act collectively. To {model the collective forces, we applied} the Dicke model~\cite{Gross1982} to each sub-domain. In this context, the $N_i$ NVs contained in the sub-domain $i$ can be represented as a superposition of $N_i+1$ spin 1/2 states. The collective force can then be obtained by solving the Liouville equations for the collective spin operator $S_{+}=\sum_{j=1}^{N_i}\left| e_j \right> \left< g_j \right|$ where $\left| e_j \right>$ and $\left| g_j \right>$ are the excited and ground state of NV number $j$ respectively. In this simplified model, the single-spin dephasing rate is given by the spontaneous decay rate of the NV, and we used a collective spin dephasing rate at room temperature of $2\pi \times1 \, \textnormal{THz}$~\cite{Fu2009} (see Supplementary Material). The collective force calculated for our particular ND sample is presented in Figure~\ref{fig3}a {as the solid black line} along with the experimental {data points}. The variability of the NV density was reproduced using a Monte Carlo approach, providing a confidence interval {within which} $90\%$ of the experimental values should fall (shaded area in Fig.~\ref{fig3}a). The number of sub-domains, and consequently their frequency width, has been used as the only adjustable parameter in our model and provides a number for NVs acting collectively. The measurements were best reproduced using a sub-domain size of approximately 100 GHz {corresponding to an average domain size of 95 NV centers, or 1$\%$ of the total average number of centers} (see Supplementary Material). Comparing with the force obtained for independent NVs, the overall impact of the collective effects is clearly apparent. We estimate an enhancement of the trap stiffness of around a factor 50, demonstrating the importance of collective effects in our analysis. In parallel {to the average trap stiffness}, we also modelled the expected skewness of the experimental data points (Fig.~\ref{fig3}b upper panel). {Our model gives} a good {quantitative} agreement with the skewness obtained from the experiment. As the model does not include any type of experimental noise, the skewness is consequently slightly over-estimated. For comparison, the skewness of the experimental data obtained {for the reference ND sample is} presented in the lower panel of Fig.~\ref{fig3}b.

In conclusion, our observations open the door to a wealth of new research
directions. The collective effects arising from the high number of NV centers in an individual nanocrystal provide a mechanism to significantly increase the optical forces. While our simplified model is in good agreement with the experiment, a more complete description accounting for dipole-dipole interactions~\cite{Gross1982} constitutes an interesting future research direction. In addition, the observed $10\%$ change in trapping stiffness could be further increased by using defects such as silicon-vacancy centers which are characterized by higher densities~\cite{Vlasov2014} and {stronger transition dipole moment}~\cite{Rogers2014,Sipahigil2014}. These centers could offer the opportunity to {access a regime in which} the resonant trapping forces dominate the dynamics of the system, with the nanocrystal essentially behaving like a very large atom, or superatom. In the context of quantum opto-mechanics, this could allow for single-photon strong coupling and side-band cooling at room temperature~\cite{juan2015}. {With all these exciting possibilities at hand, this work opens the door to applying the} powerful quantum technologies developed for atom trapping and 
cooling to the manipulation of small nanoparticles introducing an unprecedented degree of control at the nanoscale.

\section*{Acknowledgments}
We thank O. Romero-Isart for useful discussions. This work was
funded by the Australian Research Council Centre of Excellence for
Engineered Quantum Systems (EQuS) CE 110001013. G.M.-T. acknowledges funding by
the Australian Research Council Future Fellowship program. Comments or requests for materials should be addressed to \mbox{mathieu.juan@mq.edu.au} or \mbox{thomas.volz@mq.edu.au}.

\section*{Materials and Methods}

\paragraph{Nanodiamond (ND) Sample} The {two nanodiamond samples used in this
work are synthetic type Ib diamond powders with a nitrogen
concentration of 300 ppm, i.e. $3\times10^7$ nitrogen atoms per
$\mu$m$^3$ (MSY $\textless$0.1 $\mu$m; Microdiamant). This ND powder
was only chemically and mechanically processed to remove the sp$^2$
carbon-phase in excess \cite{Bradac2009}. The NDs with no additional treatment were used as reference (low NV centers concentration)  in the
control experiment to determine the effect of chromatic aberrations
in the standing wave trap. For high NV centers concentration,} the ND powder was further treated to increase the
concentration of NV centres (Academia Sinica, Taipei Taiwan) as
follows: the nanodiamonds were purified by nitration in concentrated
sulphuric and nitric acid (H$_2$SO$_4$-HNO$_3$), rinsed in deionized
water, irradiated by a 3-MeV proton beam at a dose of $1\times10^6$
ions per cm$^2$ and annealed in vacuum at 700 degrees Celsius for 2 hours to
induce the formation of NV centres \cite{Fu2007}. Prior to the
trapping experiment, both nanodiamond samples were characterized by
means of a lab-built confocal scanning fluorescence microscope (100x
oil immersion objective UplanFL N, NA 1.3; Olympus) excited with a
532-nm CW diode-pumped solid-state laser (Compass 315-M100; Coherent
Scientific) and combined with a commercial atomic force microscope
(Ntegra; NT-MDT) \cite{Bradac2009}. For characterization, the
diamond nanocrystals were dispersed on 170-$\mu$m thick BK7 glass
coverslips (BB022022A1; Menzel-Glaser) which were previously
sonicated and rinsed in acetone (C3H6O, purity $\ge$ 99.5\%;
Sigma-Aldrich) for 10 min. The measured average size of the
nanodiamonds is (150.5 $\pm$ 23.3) nm,
determined by atomic force microscopy and confirmed by dynamic light
scattering analysis (Zetasizer Nano-ZS; Malvern Instruments). The
spectral interrogation of the NDs to identify emission from NV$^{-}$
centres was performed via a commercial spectrometer (Acton 2500i,
Camera Pixis100 model 7515-0001; Princeton Instruments). While for
the untreated sample the concentration of NV centres is extremely
low (at most a few NVs per nanocrystals), for the irradiated one we
estimate $\sim 10^4$ NV centres per nanodiamond. This
was determined by correlating, for nanocrystals of different sizes,
the average fluorescence intensity measured for each ND with its
volume, and comparing this ratio with the one given by the sample
provider \cite{Fu2007}.

\paragraph{Trapping Setup} For the trapping experiment, the suspension of
nanodiamonds in deionized water was inserted in a microfluidic
chamber consisting of a BK7 glass coverslip (BB022022A1;
Menzel-Glaser) and a protected silver mirror (PF10-03-P01; Thorlabs)
using double sided tape for sealing (50 mm $\times$ 50 m, 0.14
mm-thick; 3M). The experiment involves five diode lasers, four used
for the optical trapping itself and one for the re-pumping from
NV$^{0}$ to NV$^{-}$. The re-pump laser is a 532-nm pulsed laser
(LDH-P-FA-530B; PicoQuant) used at 40 MHz repetition rate with an
average output power at the sample of 30 $\mu$W. The four other lasers are
temperature-stabilized laser diodes combined through the same fibre
to ensure that the focal spots of the different diodes are
superimposed perfectly in lateral direction. One of these diodes is
used to provide a stable conventional trap for the ND at a
wavelength of 660 nm with 6~mW of power (660 nm/130 mW; Oclaro) in order to maintain
the ND trapped during the measurement. The three remaining diodes provide the
reference wavelength ($\lambda_{ref}$), the blue and the red
wavelengths with 4~mW of power. The choice of diodes (Oclaro 633 nm/110 mW, Oclaro 637
nm/170 mW, Mitsubishi 638 nm/150 mW and Oclaro 642 nm/150 mW)
provides an overall covered spectral range of 629-648 nm (see
Extended Data Fig. 2a). The laser beams are switched using
home-built electromechanical shutters controlled with a data
acquisition system (NI-PCI 6289; National Instrument) in order to
provide 50-s continuous time trace composed of 10-s segments with
different wavelengths (see Extended Data Fig. 2b). The trapping beam
at the output of the fiber is polarized using a Glan-Laser calcite
polarizer (GL10-A; Thorlabs) and then focused into the chamber
through a water-immersion objective (UPLSAPO 60XW, NA 1.2; Olympus) with a measured waist at 640~nm of $w_0$=470~nm.
The position of the objective is set such that the reference laser
diode (639.13~nm) is focussed on the mirror forming the top surface
of the microfluidic chamber. Finally, the wavelength for the
reference laser was chosen very close to the measured average NV ZPL
position of 639.08 nm (see Extended Data Fig. 1b).

\paragraph{Stiffness measurement} Using a quadrant photodiode, the position of the ND is tracked over time. A Fourier transform of the signal yields the corresponding power spectral density
(PSD) in reciprocal space. From a Lorentzian fit to the PSD
\cite{Neuman04}, the corner frequency $f_c$ is extracted which can
be directly related to the trap stiffness through $\kappa = 2\pi
\beta f_c$, where $\beta$ is the drag coefficient of the ND. In the experiment, we measure $f_c$ as a
function of wavelength in the vicinity of the ZPL of the NV$^-$. The drag coefficient $\beta$ is a function of the viscosity of the medium, the ND size and its distance to the surface~\cite{Neuman04}. Consequently, referencing every measurement to a reference wavelength $f_c(\lambda_{measure})/f_c(\lambda_{ref})$ allows direct access to the ratio of the stiffnesses without explicit knowledge of the drag coefficient. In addition, we subtracted to each measurements ($\lambda_{red}$+660-nm, $\lambda_{blue}$+660-nm or $\lambda_{ref}$+660-nm) the corner frequency obtained for the 660-nm laser in order to access the stiffness ratio for $\lambda_{red}$, $\lambda_{blue}$ or $\lambda_{ref}$ (see Supplementary Information).


\pagebreak
\newpage
\setcounter{figure}{0}


\begin{figure}
    \includegraphics[scale=0.85]{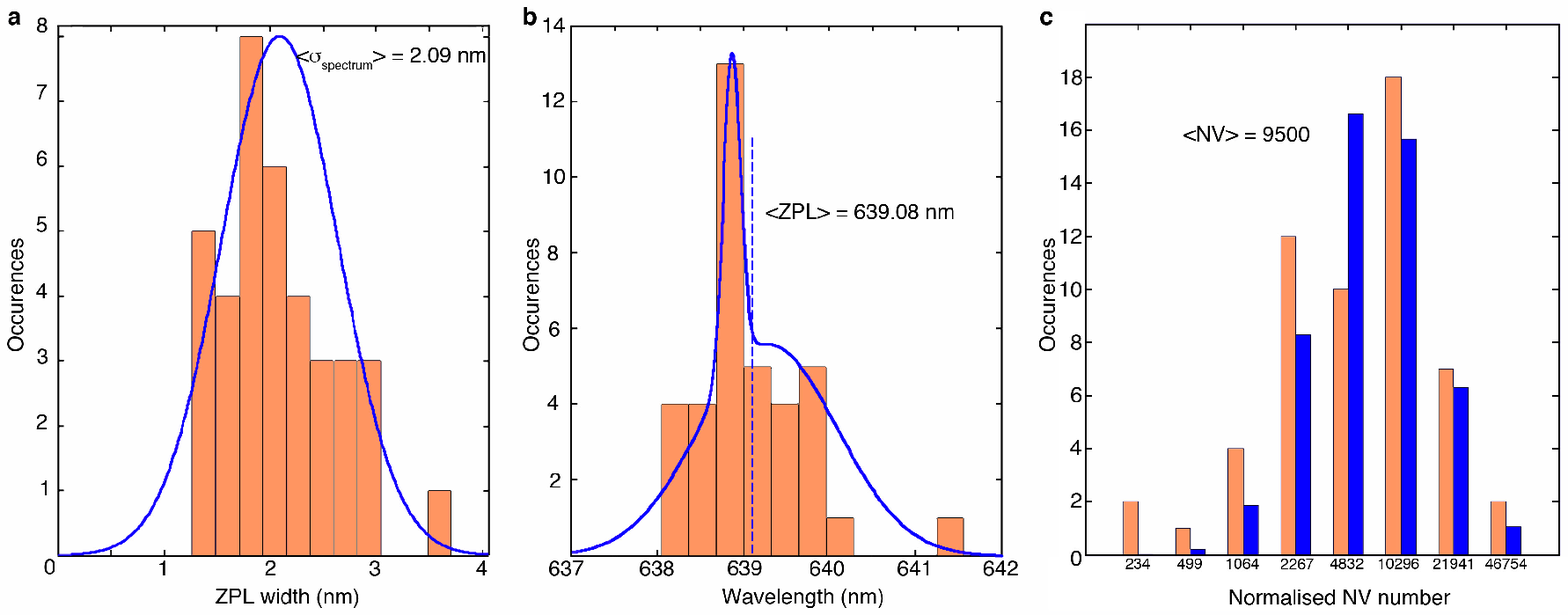}
\centering
\caption{\textbf{Extended Data Fig. 1. ND properties.} All the
properties are extracted from photoluminescence (PL) spectra
obtained on 40 different NDs. \textbf{a,} Spectral width of the ZPL
peak. The distribution was fitted by a normal distribution (in
blue), giving an average value of $\sigma_{spectrum} =
2.09\,\mathrm{nm}$. \textbf{b,} Position of the ZPL peak. The
distribution consists of two clearly distinct populations which we
approximated by two normal distributions. The overall average ZPL
position is $\left< ZPL \right> = 639.08\,\mathrm{nm}$. \textbf{c,}
Estimated number of NVs normalized to a $100\,\mathrm{nm}$ ND. The
actual values were obtained by referencing the observed average photo-luminescence
counts to the values obtained by the sample provider in
Ref~\cite{Fu2007}. \textbf{a, b, c, } the experimental data is presented in orange and the model in blue.}
\end{figure}

\begin{figure}
    \includegraphics[scale=1]{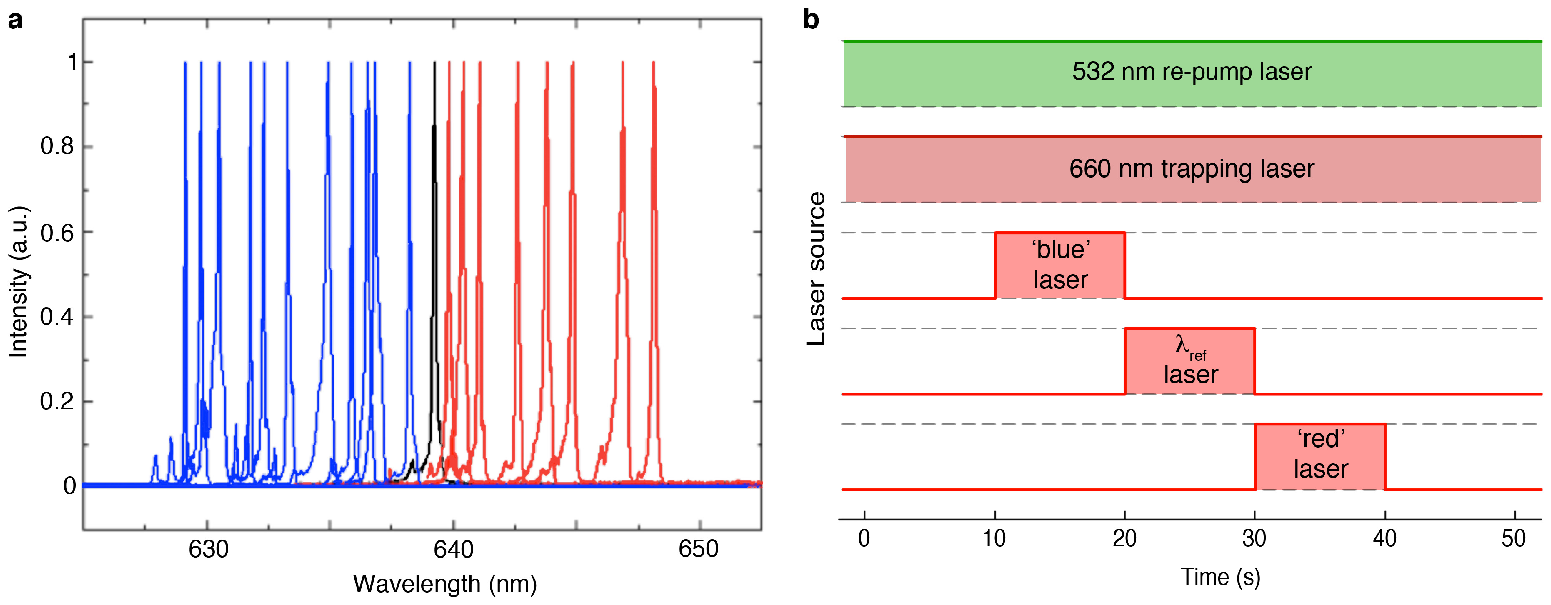}
\centering
\caption{\textbf{Extended Data Fig. 2. Laser sources and timing
sequence.} \textbf{a,} Spectra of the different laser diodes used in
the experiment: shown in black is the output from the reference laser
diode ($\lambda_{ref}$), in blue the output from the ``blue" laser
diodes ($\lambda < \lambda_{ref}$) and in red the output from the
``red" laser diodes ($\lambda > \lambda_{ref}$). \textbf{b,} Timing
sequence used during the 50-s acquisition interval. The 660-nm and
532-nm lasers are kept on all the time: the 660-nm laser provides
efficient trapping, the 532-nm laser re-pumps NV centers from the
NV$^0$ to the NV$^-$ state. The first and last 10 seconds of the
acquisition window give the trap stiffness for the 660-nm trapping
laser only. In the intermediate 30 seconds, there are three 10-s
acquisition windows for the ``blue", reference and ``red" lasers,
respectively.}
\end{figure}

\pagebreak
\clearpage
\widetext
\begin{center}
\textbf{\large Supplemental Materials: Observation of cooperatively enhanced atomic dipole forces from NV centers in optically trapped nanodiamonds}
\end{center}
\setcounter{equation}{0}
\setcounter{figure}{0}
\setcounter{table}{0}
\setcounter{page}{1}
\makeatletter
\renewcommand{\theequation}{S\arabic{equation}}
\renewcommand{\thefigure}{S\arabic{figure}}
\renewcommand{\bibnumfmt}[1]{[S#1]}
\renewcommand{\citenumfont}[1]{S#1}


\subsection*{Data acquisition and treatment}

\subsubsection{Data acquisition and sequencing} The data
was obtained from an acquisition time trace of 50~s overall duration.
The timing sequence for applying the different trapping lasers
during these 50~s is displayed in the Extended Data Fig. 2. The time
trace is partitioned in equal segments of 10 seconds, and the corner
frequency during these 10 seconds is obtained from the power
spectral density of each individual segment. During the first and
last 10~s of acquisition, only the 660-nm laser is on corresponding
to the corner frequency $f_{660}$. During each of the other 10~s
intervals, the 660-nm laser is on plus one of the three following
lasers at a time: the ``blue" laser with $\lambda < \lambda_{ref}$
(giving the corner frequency $f_{blue}$), the reference laser with
$\lambda = \lambda_{ref}$ ($f_{ref}$), and the ``red" laser with
$\lambda > \lambda_{ref}$ ($f_{red}$). Since the 660-nm laser is far detuned from the resonance the dipole force from the NV transition with this laser can be neglected. This point stands when using both the 660-nm laser and a laser closer to the NV transition (e.g. ``red" laser): the dipole force from the NV due to the ``red" laser does not depend on the 660-nm laser (for moderate power as such used in this work). Within the harmonic approximation, the overall corner frequency with two lasers on
simultaneously (e.g. the 660-nm laser and the laser with
$\lambda_{ref}$) can thus be written as: $f_{ref} =
(\kappa_{660}+\kappa_{ref})/(2\pi\beta)$ with $\beta$ the drag
coefficient of the particular nanodiamond. The observables of
interest, $\kappa_{blue}/\kappa_{ref}$ and
$\kappa_{red}/\kappa_{ref}$, are therefore determined by
\begin{equation}
\frac{\kappa_{blue}}{\kappa_{ref}} = \frac{f_{blue} -
f_{660}}{f_{ref} - f_{660}} \ \ \mathrm{and} \ \
\frac{\kappa_{red}}{\kappa_{ref}} = \frac{f_{red} - f_{660}}{f_{ref}
- f_{660}}
\end{equation}
Normalizing the corner frequencies this way gives access to the
trapping stiffness ratios without requiring knowledge of the actual size of
the trapped NDs. In addition, the normalization avoids the
dependence of the trapping stiffness on the volume.
Hence, the normalized ratios can directly be compared for different
NDs.

\subsubsection{Data treatment}
As explained previously, the first and last 10~s of the acquisition
are necessary to extract the effect of the 660-nm trapping laser
only. In addition, we compared the corner frequency for these two
segments with the 660-nm trapping laser to identify anomalous
events during the whole 50~s data acquisition. As a boundary
condition for discarding anomalous time traces, we imposed a
relative change in corner frequency smaller than $10\%$ between the
two segments. Based on this
$10\%$-rule, we typically removed $20\%$ of the data. The main
cause for a dramatic change in 660-nm corner frequency at the
beginning and end of the 50~s acquisition period would most likely be
a second ND hopping into the trap due to the relatively high
concentration of NDs in the solution.

The 10\% selection rule was applied before calculating the trapping
stiffness ratios and allowed us to discard clearly anomalous time
traces. In addition, we also used a Local Outlier Factor (LOF)
method \cite{SMBreunig2000} to remove clear statistical outliers from
the data. This method is based on calculating the local density of
neighbors for each data point. Within the LOF method, (local)
outliers are identified by their large LOF value. We rejected data
points having a LOF larger than 5.7 using a LOF calculation based on
the 6th-nearest neighbor. The same parameters were used for all the
data and ultimately removed 3.4\% of the remaining data points. We
verified that this method would not impact the effect we wanted to
underline. To do so, we used a Monte Carlo simulation to numerically
reproduce the experiment accounting only for the variability of the
ND properties (number of NVs, average ZPL transition and ZPL standard deviation).
Applying the very same LOF selection rule to the numerical results
only removed an average of 0.6\% of the numerical data points. This confirms that even with the large variability in the NV density, the outlier method only marginally impacts the data.

\subsection*{Calculating the dipole force for an ensemble of NVs}

\subsubsection{Dipole force on a 2-level system}

For a simple 2-level system with a dipole moment $d_{2l}$ and transition $\omega_{2l}$ in the presence of dephasing ($\gamma = \Gamma/2 + \gamma_c$) in an electric field with amplitude $E_0(\mathbf{r})$, the optical dipole potential can be written as~\cite{SMGrynberg2010}:
\begin{equation}
U_{2l}(\omega,\mathbf{r}) = - \frac{\hbar\left( \omega - \omega_{2l} \right)}{2} \frac{\Gamma}{2\gamma} \text{log}\left[ 1+s_{2l}\left(\omega,\mathbf{r}\right) \right],
\end{equation}
where $s_{2l}(\omega,\mathbf{r})=\Omega(\mathbf{r})^2/[\Gamma \gamma(1+(\omega-\omega_{2l})^2/\gamma^2)]$ is the saturation parameter and $\Omega(\mathbf{r}) = \sqrt{2/3} \, d_{2l} \, E_0(\mathbf{r})/\hbar$ the Rabi frequency. Here the prefactor $\sqrt{2/3}$ comes from the time average of the electric field ($\sqrt{2}$) and the orientational average of the dipole moment ($1/\sqrt{3}$).

\subsubsection{NV center and phonon sidebands}

NV centers constitute good quantum emitters, yet their level structure is quite different from the ideal 2-level system usually considered in calculating the dipole forces. In particular, the large phonon sideband (PSB) needs to be taken into account when estimating the dipolar moment. From the spontaneous rate emission in vacuum of NVs, $\Gamma_0$, at the transition frequency $\omega_0$, the overall dipolar moment $d$ can be determined via the Fermi Golden rule as $d=\sqrt{\Gamma_0 3\pi \epsilon_0 c^3 \hbar/\omega_0^3}$. Here $\epsilon_0$ is the vacuum permittivity, $c$ the speed of light and $\Gamma=n_1 \Gamma_0$ with $n_1$ the refractive index of diamond and $\Gamma$ the spontaneous rate emission of NVs in nanodiamonds~\cite{SMInam2011}. Decomposiing the PSB in 7 distinct bands following~\cite{SMAlbrecht2013}, the spontaneous rate emission can be written as $\Gamma = \Gamma_{ZPL} + \sum_{i=1}^7\Gamma_i$ with $\Gamma_{ZPL}$ the zero phonon line (ZPL) spontaneous rate emission and $\sum_{i=1}^7\Gamma_i = \Gamma_{SB}$ the sum on the phonon sidebands~\cite{SMAlbrecht2013}. When studying the forces around the ZPL at $\omega_0$, the different sidebands ($\omega_i$) are far detuned enough to be ignored (min$_i\left| \omega_0 - \omega_i \right| > 10$~nm). As a consequence, we used $d_{ZPL} = d \cdot \sqrt{\Gamma_{ZPL}/\Gamma}$ for the dipole moment:
\begin{equation}
\label{eq:fermi} d_{ZPL} = \sqrt{\frac{\Gamma_{ZPL} 3\pi \epsilon_0 c^3 \hbar}{n_1 \omega_0^3}}
\end{equation}
Consequently, due to the phonon sidebands the dipole moment is effectively reduced. Also we used a branching ratio, or Debye-Waller factor, of 0.04~\cite{SMSiyushev2009,Faraon2011}.

In addition to reducing the dipole moment, we also took into account the sidebands calculating the dipole force. We made the following approximations in order to maintain a simplified and analytical approach: (i) the different sidebands have been accounted for as one extra channel with a decay rate going as the sum of the various bands, $\Gamma_{SB} = \sum_{i=1}^7\Gamma_i$, (ii) the non-radiative phonon decay rate is assumed to be similar for all bands and equal to $\Gamma_{Ph}/(2\pi) =38 \textnormal{GHz}$~\cite{SMHuxter2013}, and (iii) the sidebands cannot be efficiently coherently driven due their very large dephasing ($\gtrsim 60\textnormal{THz}$) and large detuning from the laser resonant with the ZPL. Under these approximations, the relevant terms from the optical Bloch equation can be explicitly written as:
\begin{eqnarray} \nonumber
\frac{\partial }{\partial t}\rho_{ee} & = &  i \frac{\Omega}{2} \left( \tilde{\rho}_{eg} - \tilde{\rho}_{ge} \right)  - \left(\Gamma_{ZPL} + \Gamma_{SB} \right) \, \rho_{ee} \\
\frac{\partial}{\partial t}\rho_{gg}  & = & -i \frac{\Omega}{2} \left( \tilde{\rho}_{eg} - \tilde{\rho}_{ge} \right) + \Gamma_{ZPL} \, \rho_{ee} + \Gamma_{Ph} \, \rho_{pp} \\  \nonumber
\frac{\partial }{\partial t}\rho_{pp} & = & \Gamma_{SB} \, \rho_{ee} - \Gamma_{Ph} \, \rho_{pp} \\  \nonumber
\frac{\partial }{\partial t}\tilde{\rho}_{eg} & = & - \left( \gamma -i\Delta\right) \tilde{\rho}_{eg} +i \frac{\Omega}{2} \left( \rho_{ee} - \rho_{gg} \right),
\end{eqnarray}
where $g$, $e$ and $p$ are the ground state, the excited state and the phonon sideband respectively, $\Omega(\mathbf{r}) = \sqrt{2/3} \, d_{ZPL} \, E_0(\mathbf{r})/\hbar$ is the Rabi frequency and $\Delta$ the laser detuning from the transition $\Delta = \omega -\omega_0$. The density matrix element $\tilde{\rho}$ is the element $\rho$ in the rotating frame, $\tilde{\rho} = e^{-i \omega_0 t}\rho$. The decay rate from the excited state to the phonon sideband $e \rightarrow p$ is $\Gamma_{SB}$, from the excited state to the ground state $e \rightarrow g$ is $\Gamma_{ZPL}$, and from the phonon sideband to the ground state $p \rightarrow g$ is $\Gamma_{Ph}$. At room temperature, the total linedwith of the NV is much larger than the lifetime limited linewidth. This linewidth (or transverse decay rate) is $\gamma = (\Gamma_{ZPL} + \Gamma_{SB})/2 + \gamma_c$, where $\gamma_c$ accounts for additional coherence decay (inhomogeneous broadening). The value typically observed at room temperature is $\gamma/(2\pi)\sim1$~THz~\cite{SMFu2009}.

In the Heisenberg-picture, the time averaged value of the dipole force can then be written as:
\begin{equation}
\left< \mathbf{F}_{dip} \right>  = -\frac{\hbar \Omega^{\ast} \left(\mathbf{r} \right) \left< \sigma \right>}{2}   \nabla \textnormal{log}\left| \Omega\left(\mathbf{r} \right) \right|+ \textnormal{c.c},
\end{equation}
where $ \left< \sigma \right>= \tilde{\rho}_{eg}$ is the steady state coherence as obtained by solving the steady state optical Bloch equation for $\tilde{\rho}_{eg}$, and $\Omega^{\ast}$ is the complex conjugate of $\Omega$. The force can then be explicitly written as:
\begin{equation}
\mathbf{F}_{dip}\left(\mathbf{r} \right) = - \frac{1}{\eta} \frac{\hbar \Delta}{2} \frac{\Gamma}{2\gamma} \frac{\nabla s\left(\mathbf{r} \right)}{1+s\left(\mathbf{r} \right)},
\end{equation}
with $\eta=(2\Gamma_{Ph}+\Gamma_{SB})/2\Gamma_{Ph}$ and the saturation parameter $s\left(\mathbf{r} \right)=\eta \Omega\left(\mathbf{r} \right)^2/[\Gamma \gamma(1+\Delta^2/\gamma^2)]$. The impact of the phonon sidebands contained in the factor $\eta$ is clearly negligible if the phonon decay rate $\Gamma_{Ph}$ (phonon sideband $\rightarrow$ ground) is the fastest decay. Intuitively, when this decay is fast the phonon sidebands are never populated (i.e. $\rho_{pp}=\rho_{ee} \Gamma_{SB}/\Gamma_{Ph}  \ll \rho_{ee}$) and the system behaves as a typical two-level system obtained for $\eta = 1$. Conversely, if this decay is slow enough to maintain population in the phonon sidebands both the saturation parameter and the force are impacted. In terms of optical potential one obtains:
\begin{equation}
U_q\left(\omega,\omega_0,\mathbf{r} \right) = - \frac{1}{\eta} \frac{\hbar \left(\omega-\omega_0\right)}{2} \frac{\Gamma}{2\gamma} \textnormal{log}\left[1+s\left(\mathbf{r} \right)\right]
\end{equation}

\subsection*{ Trapping stiffness}

The fitting of the experimental data is done on the trapping stiffness rather than the optical potential depth. For this reason, the position dependence of the potential has to be given explicitly. Along the measurement direction $x$, we approximated the electromagnetic field to a Gaussian profile $E(x) = E_0 \textnormal{exp}(-x^2/w_0^2)$, with the field amplitude $E_0 = \sqrt{4P/(\pi w_0^2 n_2 \epsilon_0 c)}$. Here the waist of the focus is defined by $w_0$ ($w_0$=470~nm, measured at 640~nm), the incident power by $P$ and the refractive index of the medium (water) by $n_2$.

The optical potential for the force acting on the nanodiamond dielectric matrix is then given by~\cite{SMHarada1995}:
\begin{equation}
\label{eq:Cl_pot} U_{cl}(x) = - {2\pi \epsilon_0 n_2^2 R^3}\left(\frac{m^2-1}{m^2+2} \right)E_0^2 \mathrm{exp}{\left[-\frac{2x^2}{w_0^2} \right]},
\end{equation}
where $m=n_1/n_2$ is the refractive index of diamond ($n_1$) relative to the medium (water, $n_2$). Similarly, the optical potential acting on the NV center can be explicitly given by:
\begin{equation}
\label{eq:Q_pot} U_q(x) = \frac{\hbar \left(\omega - \omega_0\right)}{2 \eta} \frac{\Gamma}{2\gamma} \textnormal{log}\left[1 + \eta \frac{2 d_{ZPL}^2 E_0^2}{3\hbar^2} \frac{1/(\Gamma \gamma)}{1 + (\omega -\omega_0)^2/\gamma^2}\mathrm{exp}{\left(-\frac{2x^2}{w_0^2} \right)} \right]
\end{equation}
For the sake of simplicity, in the following we refer to the first
potential (force) as the ``classical" potential (force) and to the
second one as ``quantum" potential (force).

In the overdamped regime, the variance in the position of the
trapped ND can be obtained through the equipartition theorem $\Delta
x= \sqrt{k_BT\left(2\pi f_c \beta\right)^{-1}}$, with $\beta$ being
the drag coefficient of the ND. This variance is estimated to be
around $40$~nm for a $150$~nm ND and an average corner frequency of
$\ f_c \approx 400$~Hz. Such a small displacement allows
neglecting any anharmonic effects such as for example a Duffing
non-linearity \cite{SMGieseler2013}. In this case the optical
potential seen by the ND can then be well approximated by a simple
harmonic potential of the form $U \sim U(0)+ \kappa x^2$ where
$\kappa$ is related to the potential through $\kappa =\frac{1}{2}
\partial^2 U(x)/\partial x^2|_{x=0}$. For the simple case of a nanodiamond containing $N$ identical NVs with transition frequencies $\omega_0$, the two components of the stiffness
can thus be written as:
\begin{equation}
\label{eq:kappa_cl} \kappa_{cl} = \frac{4 \pi \epsilon_0 n_2 R^3}{w_0^2}\left(
\frac{m^2-1}{m^2+2} \right)E_0^2
\end{equation}
\begin{equation}
\label{eq:kappa_q} \kappa_{q} =- N \frac{\hbar \left(\omega - \omega_0
\right)}{2 \eta} \frac{\Gamma}{2 \gamma} \frac{4}{w_0^2}
\left( \frac{s_0(\omega_0,\omega)}{1+s_0(\omega_0,\omega)} \right),
\end{equation}
with  $s_0(\omega_0,\omega)$ the saturation parameter taken at $x=0$ given by:
\begin{equation}
 s_0(\omega_0,\omega)= \frac{\eta}{\Gamma \gamma} \frac{2 d_{ZPL}^2 E_0^2}{3 \hbar^2} \frac{1}{1+(\omega-\omega_0)^2/\gamma^2}
\end{equation}
Note that the waist of the Gaussian beam, $w_0$, is also a
(monotonic) function of wavelength.

\subsection*{Cooperative effects}

When calculating the dipole force from an ensemble of NVs such as the one present in the nanodiamonds used in the experiment, this force is much smaller than the one observed experimentally. Due to the unusually high density of NVs in these samples, it is necessary to account for cooperative effects in order to fully capture the dipole force from the NV ensemble. In addition, the small size of the nanodiamonds ($150$~nm) allows for a simple description without finite size effects. Consequently, we used a simple model based on the Dicke model~\cite{SMGross1982}.

Due to the dephazing and the slightly different transition frequency of each NV center, only sub-populations of NVs are expected to act cooperatively. This amounts to ``coarse-grain" the ensemble of $N$ NV centers into cooperative sub-ensembles. Within one such sub-ensemble $i$ containing $N^i_{Coop}$ NVs, we assume the system to be invariant under permutation in order to apply the Dicke model. In this context, the state space in each sub-domain is spanned by Dicke states $\left| J,M\right>$ with $J=N^i_{Coop}/2$. The maximal angular momentum space is appropriate because the spin begin in the ground state $\left| J, -J \right>$. We also assume that the PSB does not impact the excited and ground populations (\textit{i.e.} $\eta \approx 1$). The raising and lowering operators for the NV defect $k$ are defined respectively as:
\begin{equation}
S_k^{+} 	\equiv \left| e \right> \left< g \right| \textnormal{;} \, \, \, \, \, \, \, \, \, \, \, \, \, S_k^{-} 	\equiv \left| g \right> \left< e \right|
\end{equation}
end the diagonal operator as:
\begin{equation}
S_k^z \equiv \frac{1}{2} \left[ \left| e \right> \left< e \right| - \left| g \right> \left< g \right|\right]
\end{equation}
For notational clarity we suppress the sub-domain label $i$ on the collective spin operators but it is to be understood that the collective spin operators appearing in the equations to follow for the forces and stiffness ratios carry such an index.

Let us now consider one sub-domain $i$ containing $N_{Coop}^i$ NVs with a transition $\omega_i$. Using the collective operators for the collective sub-domain $i$, $S^{\pm} \equiv \sum_{k=1}^{N_{Coop}^i} S_k^{\pm}$ and $S^z \equiv  \sum_{k=1}^{N_{Coop}^i} S_k^z$, the symmetrical state can be obtained by repeated action of the lowering operator on the fully excited state:
\begin{equation}
\left| JM \right> = \sqrt{\frac{\left( J + M \right) !}{N_{Coop}^i ! \left( J - M \right) !}} \cdot \left(S^- \right)^{J-M} \left| e,\,e,\,e \ldots e \right> 
\end{equation}
with $-J \leq M \leq J = N_{Coop}^i/2$. 
\\ This $\left| J M \right>$ state represents the fully symmetrical collective state with $J+M$ atoms in the excited state $\left| e\right>$ and $J-M$ in the ground state $\left| g\right>$. The Hilbert space dimension of this collective spin state is then $N_{Coop}^i + 1$.

We then consider this system to be driven by a laser at frequency $\omega$ according to the Hamiltonian $H$ given, in the the rotating wave approximation, by:
\begin{equation}
H = - \left( \omega - \omega_i \right) S^z + \frac{\Omega}{2} \left( S^+ + S^- \right)
\end{equation}
where $\omega_0$ is the spin transition frequency and $\Omega$ the Rabi frequency. The Louivillian describing the coupling of this system to the laser and the environment is then:
\begin{equation}
\begin{aligned}
\dot{\rho}\left( t \right) = & -\frac{i}{\hbar} \left[ H, \rho (t) \right]\\
 &- \frac{\gamma_{bare}}{2} (\{S^+S^-,\rho(t)\}-2 S^-\rho(t) S^+)+\frac{\gamma_{collective}}{2}(\{S^zS^z,\rho(t)\}-2 S^z\rho(t) S^z)
 \label{dynamics}
\end{aligned}
\end{equation}
where the decay rate $\gamma_{bare}$ is the bare single spin decay rate, and $\gamma_{collective}$ is a collective dephasing rate. Since the observed inhomogeneous broadening of NV centers at room temperature has been attributed to phonon processes~\cite{SMFu2009}, this dephasing mechanism is believed to act on the collective state as it will affect all spins indiscernibly. As a consequence, we used the spontaneous decay rate of the NVs, $\Gamma$, as the single spin decay rate and inhomogeneous broadening, $\gamma_c$ as the collective dephasing. In this context, the dipole force is given by:
\begin{equation}
\mathbf{F} = - \Re \left[ \hbar \mathbf{\nabla} \Omega \, \left< \Sigma^+ \right> \right]
\end{equation}
where $\left< \Sigma^+ \right> = \textnormal{Tr} \left[ S^+ \rho^{SS} \right]$ is the expectation value of $S^+$ in the steady state of the collective spin state $\rho^{SS} = \textnormal{lim}_{t\rightarrow \infty} \rho(t)$.

We can numerically compute the reactive force by finding the steady state of the collection of spins as follows. First we vectorize Eq. \ref{dynamics} by writing the density operator as a single vector of dimension $(N_{Coop}^i+1)^2$:
$\rho\rightarrow \vec{\rho}$, which is done by stacking the rows of $\rho$ on top of one another (such that the state $\ket{0}$ has a one in the first component and zeros otherwise). The dynamical equation is then
\[
\dot{\vec{\rho}}(t)=A \vec{\rho}(t)
\]
where the Louvillian in the vectorized basis is
\begin{equation}
\begin{array}{lll}
A&=&i (\omega-\omega_i) (S^z\otimes \mathds{1}- \mathds{1}\otimes S^z) +i\frac{\Omega}{2} ((S^-+S^+)\otimes \mathds{1}- \mathds{1}\otimes (S^-+S^+))\\
&&-\frac{\Gamma}{2} (S^+S^-\otimes \mathds{1}+\mathds{1}\otimes S^+S^- -2 S^-\otimes S^-) \\
&& -\frac{\gamma_c}{2}((S^z)^2\otimes \mathds{1}+\mathds{1}\otimes (S^z)^2-2 S^z\otimes S^z).
\end{array}
\end{equation}
where $\mathds{1}$ is the unit vector of dimension $N_{Coop}^i+1$.
The steady state is contained in the null space of $A$ which for this case is one dimensional since the dynamical algebra is irreducible.

We also verified that the positional variance of the nano-diamond in the trap, $\Delta x \approx 40\textnormal{ nm}$, does not influence the steady-state population. Under the harmonic trap approximation, the stiffness along $x$ for the sub-domain $i$ can be obtained numerically as:
\begin{equation}
\kappa_i(\omega, \omega_i,N_{Coop}^i) =- \hbar \left. \frac{\partial^2 \Omega}{\partial x^2}\right|_{x=0} \, \Re \left[ \left< \Sigma^+ \right> \right]
\end{equation}
Using this method, we calculated the total ``quantum" stiffness $\kappa_{q}$ as the sum of the stiffness $\kappa_i$ from the different collective domains: 
\begin{equation}
\kappa_q \left( \omega \right)= \sum_{i=-N_{Grains}/2}^{N_{Grains}/2} \kappa_i\left(\omega,\omega_i, N^i_{Coop}\right)
\end{equation}
As the size of the vectorized steady state is $(N_{Coop}^i+1)^2$, the numerical solution of the steady state populations becomes difficult for large sub domain size. Consequently, we calculated the force obtained with $N_{Coop} \in \left[ 1, 80\right]$ and extrapolated the force for larger sub-domain using polynomial fits.

Note the collective dephasing preserves the total angular momentum because the environment that induces it has interactions which are permutation-symmetric. Local dephasing would couple the system out of this fixed angular momentum subspace, reducing the number of spins in the collective sate over time. In the context of this work focusing on the force, the steady state of the whole system constitute the most important aspect. The underlying assumption is that under continuous drive and in conjunction with local dephasing, the cooperative sub-domains will approach a steady state of fixed mean size. The size of these sub-domains was not determined theoretically, but rather used as the only fitting parameter to reproduce our experimental data.

To define the size of sub-domains, the distribution of NV transition frequencies within a given nanodiamond has been modelled as a normal distribution centered at $\omega_0$ with a standard deviation $\sigma_{ZPL}$ (see Extended Figure 1). The coarse-graining then consists in dividing this distribution in sub-ensembles of spectral width $\Gamma_{Grains}$. Within each of these ensembles $i$ ($i \in \mathbb{Z}$) the NVs are assumed to form a collective state with a transition frequency $\omega_i$ and a size $N^i_{Coop}$ defined as:
\begin{eqnarray} \nonumber
\omega^i_{start} &=& (i-\frac{1}{2}) \Gamma_{Grains}  + \omega_0\\ 
\omega^i_{end} &=& (i+\frac{1}{2}) \Gamma_{Grains} + \omega_0\\ \nonumber
N^i_{Coop} &=& \int\limits_{\omega^i_{start}}^{\omega^i_{end}} \mathrm{d}\omega \frac{N }{\sigma_{ZPL}\sqrt{2\pi}}\ \mathrm{exp}\left[
\frac{-\left(\omega_0 - \omega\right)^2}{2\sigma_{ZPL}^2}\right] \\ \nonumber
\omega_i  &=& \omega_0 +  i \Gamma_{Grains}
\end{eqnarray}
As the normal distribution of NVs in the nanodiamonds has a finite standard deviation $\sigma_{ZPL}$, only the sub-ensembles $i$ such that $\omega_i \, \in \left[\omega_0 - 4\sigma_{ZPL}, \omega_0 + 4\sigma_{ZPL} \right]$ are considered. Using this $4\sigma_{ZPL}$ interval allows one to account for $99.993\%$ of the NV population, it is also then possible to define the number of sub-ensembles $\mathcal{N}_{Grains} = 8\sigma_{ZPL}/\Gamma_{Grains}$. It is the spectral width of these sub-domains, $\Gamma_{Grains}$, or equivalently the number sub-domains  $\mathcal{N}_{Grains}$, that we used as the sole fitting parameter used to reproduce our experimental data.

\subsection*{Stiffness ratio}

In the harmonic approximation, the measurement at a given wavelength
yields $\kappa_{tot}(\lambda) = \kappa_{NVs}(\lambda) +
\kappa_{Diamond}(\lambda)$, where $\kappa_{NVs}$ is the stiffness related to the NV centers and $\kappa_{Diamond}$ the stiffness related to the diamond matrix. By normalizing with the reference wavelength,
we obtain the following ratio:
\begin{equation}
\mathrm{Ratio}(\lambda) = \frac{ \kappa_{NVs}(\lambda) +
\kappa_{Diamond}(\lambda) } { \kappa_{NVs}(\lambda_{ref}) +
\kappa_{Diamond}(\lambda_{ref}) }
\end{equation}
In the case of the low NV-density NDs, the impact of the NVs on the total force
force is neglected. The difference between the two ratios
consequently gives:
\begin{equation}
\label{eq:delta} \Xi(\lambda) =  \left( \frac{ \kappa_{NVs}(\lambda)
+ \kappa_{Diamond}(\lambda) } { \kappa_{NVs}(\lambda_{ref}) +
\kappa_{Diamond}(\lambda_{ref})} \right) _{high NV} -  \left( \frac{
\kappa_{Diamond}(\lambda) } {  \kappa_{Diamond}(\lambda_{ref}) }\right) _{low
NV}
\end{equation}

The choice of the reference wavelength $\lambda_{ref}$ at the
average ZPL was such that the contribution of the NVs on the force
is negligible (\textit{i.e.} $\kappa_{NVs}(\lambda_{ref}) +
\kappa_{Diamond}(\lambda_{ref})\approx \kappa_{Diamond}(\lambda_{ref})$). With this approximation, the difference of ratios
between high and low NV-density NDs simply yields the ratio of the
stiffness from the NV centers to the stiffness from the diamond matrix at $\lambda_{ref}$:
\begin{equation}
\Xi(\lambda) \approx \frac{\kappa_{NVs}(\lambda)}{\kappa_{Diamond}(\lambda_{ref})}
\end{equation}



\begin{thebibliography}{99}

\bibitem{Ashkin86}
Ashkin, A., Dziedzic, J.~M., Bjorkholm, J.~E., \& Chu, S.
\newblock Observation of a single-beam gradient force optical trap for dielectric particles.
\newblock \textit{Opt. Lett.} \textbf{11} 288-290 (1986).

\bibitem{Neuman04}
Neuman, K. C., \& Block, S. M.
\newblock Optical trapping.
\newblock \textit{Review of Scientific Instruments} \textbf{75}, 2787 (2004)

\bibitem{Schlosser2001}
Schlosser, N., Reymond, G., Protensko, I., \& Grangier, P.
\newblock Sub-{P}oissonian loading of single atoms in a microscopic dipole trap.
\newblock \textit{Nature} \textbf{411} 1024-1027 (2001).

\bibitem{Grier03}
Grier, D. G.
\newblock A revolution in optical manipulation
\newblock \textit{Nature} \textbf{424}, 810-816 (2003)


\bibitem{Grimm2000}
Grimm, R., Weidem{\"u}ller, M., \& Ovchinnikov, Y. B.
\newblock Optical dipole traps for neutral atoms
\newblock \textit{Advances in Atomic, Molecular and Optical Physics } \textbf{42} 95-170 (2000).

\bibitem{Grynberg2010}
Grynberg, G., Aspect, A., and Fabre, C.
\newblock \textit{Introduction to Quantum Optics From the Semi-classical Approach to Quantized Light}
\newblock Cambridge University Press, Cambridge (UK).


\bibitem{Zemanek1999}
Zem\'anek, P., Jon\'a\v s, A., \v Sr\'amek, L., \& Li\v ska, M.
\newblock Optical trapping of nanoparticles and microparticles by a Gaussian standing wave.
\newblock \textit{Opt. Lett.} \textbf{24}, 1448-1450 (1999).


\bibitem{Doherty2013}
Doherty, M.~W. \textit{et al.}
\newblock The nitrogen-vacancy colour centre in diamond.
\newblock \textit{Physics Reports} \textbf{528} 1-45 (2013).

\bibitem{Jelezko2004b}
Jelezko, F., Gaebel, T., Popa, I., Gruber, A., \& Wrachtrup, J.
\newblock Observation of Coherent Oscillations in a Single Electron Spin.
\newblock \textit{Phys. Rev. Lett.} \textbf{92}, 076401 (2004).

\bibitem{Balasubramanian2008}
G.~Balasubramanian, \textit{at al.}
\newblock Nanoscale imaging magnetometry with diamond spins under ambient conditions.
\newblock \textit{Nature} \textbf{455}, 648-651 (2008).

\bibitem{Maze2008}
Maze, J.~R. \textit{et al.}
\newblock Nanoscale magnetic sensing with an individual electronic spin in diamond.
\newblock \textit{Nature} \textbf{455}, 644-647 (2008).

\bibitem{Brouri2000}
Brouri, R. \textit{et~al.}
\newblock Photon antibunching in the fluorescence of individual color centers in diamond.
\newblock \textit{Opt. Lett.} \textbf{25}, 1294-1296 (2000).

\bibitem{Kurtsiefer2000}
Kurtsiefer, C. \textit{et~al.}
\newblock Stable solid-state source of single photons.
\newblock \textit{Phys. Rev. Lett.} \textbf{85}, 290-293 (2000).

\bibitem{Davies1976}
Davies, G., \& Hamer, M.~F.
\newblock Optical Studies of the 1.945 eV Vibronic Band in Diamond.
\newblock \textit{Proceedings of the Royal Society of London. Series A, Mathematical and Physical Sciences}
\textbf{348}, 285-298 (1976).

\bibitem{Faraon2011}
Faraon, A., Barclay, P.~E., Santori, C., Fu, K.-M.~C., \& Beausoleil, R.~G.
\newblock Resonant enhancement of the zero-phonon emission from a colour centre in a diamond cavity
\newblock \textit{Nat. Phot.} \textbf{5}, 301 (2011).

\bibitem{McGuinness2011}
McGuinness, L.~P. \textit{et al.}
\newblock Quantum measurement and orientation tracking of fluorescent nanodiamonds inside living cells.
\newblock \textit{Nat. Nano.} \textbf{6} 358-363 (2011).

\bibitem{Alhaddad2011}
Alhaddad, A. \textit{et al.}
\newblock Nanodiamond as a Vector for siRNA Delivery to Ewing Sarcoma Cells.
\newblock \textit{Small} \textbf{7} 3087-3095 (2011).

\bibitem{Horowitz2012}
Horowitz, V.~R., Alem\'an, B.~J., Christle, D.~J., Cleland, A.~N., \& Awschalom, D.~D.
\newblock Electron spin resonance of nitrogen-vacancy centers in optically trapped nanodiamonds.
\newblock \textit{PNAS} \textbf{109} 13493-13497 (2012).

\bibitem{Geiselmann2013}
Geiselmann, M. \textit{et al.}
\newblock Three-dimensional optical manipulation of a single electron spin.
\newblock \textit{Nat. Nano.} \textbf{8}, 175-179 (2013).

\bibitem{Neukirch2013}
Neukirch, L.~P., Gieseler, J., Quidant, R., Novotny, L., \& Vamivakas, A.~N.
\newblock Observation of nitrogen vacancy photoluminescence from an optically levitated nanodiamond.
\newblock \textit{Opt. Lett.} \textbf{38} 2976-2979 (2013).

\bibitem{Aslam2013}
Aslam, N., Waldherr, G., Neumann, P., Jelezko, F., \& Wrachtrup, J.
\newblock Photo-induced ionization dynamics of the nitrogen vacancy defect in diamond investigated by single-shot charge state detection.
\newblock \textit{New J. Phys.} \textbf{15} 013064 (2013).

\bibitem{Gieseler2013}
Gieseler, J., Novotny. L., \& Quidant, R.
\newblock Thermal nonlinearities in a nanomechanical oscillator.
\newblock \textit{Nat. Phys.} \textbf{9} 806-810 (2013).

\bibitem{Fu2007}
Fu, C.~C. \textit{et al.}
\newblock Characterization and application of single fluorescent nanodiamonds as cellular biomarkers.
\newblock \textit{PNAS} \textbf{104} 727-732 (2007).

\bibitem{Bradac2009}
Bradac, C., Gaebel, T., Naidoo, N., Rabeau, J.~R., \& Barnard, A.~S.
\newblock Prediction and Measurement of the Size-Dependent Stability of Fluorescence in Diamond over the Entire Nanoscale.
\newblock \textit{Nano Lett.} \textbf{9} 3555-3564 (2009).

\bibitem{Fu2009}
Fu, K.-M.~C. \textit{et al.}
\newblock Observation of the Dynamic Jahn-Teller Effect in the Excited States of Nitrogen-Vacancy Centers in Diamond.
\newblock \textit{Phys. Rev. Lett.} \textbf{103} 265404 (2009).

\bibitem{Dicke1958}
Dicke, R.
\newblock Coherence in Spontaneous Radiation Processes.
\newblock \textit{Phys. Rev.} \textbf{93}, 99 (1958).

\bibitem{Bradac2015}
Bradac, C. \textit{et al.}
\newblock Observation of room-temperature cooperative emission from diamond nanocrystals with large numbers of nitrogen vacancy centers
\newblock \textit{in preparation}

\bibitem{Gross1982}
Gross, M, \& Harroche, S.
\newblock Superradiance: an essay on the theory of collective spontaneous emission.
\newblock \textit{Physics Reports} \textbf{93}, 301-396 (1982).

\bibitem{Vlasov2014}
Vlasov, I.~I. \textit{et al.}
\newblock Molecular-sized fluorescent nanodiamonds.
\newblock \textit{Nat. Nano.} \textbf{9} 54-58 (2014).

\bibitem{Rogers2014}
Rogers, L.~J. \textit{et al.}
\newblock Electronic structure of the negatively charged silicon-vacancy center in diamond.
\newblock \textit{Phys. Rev. B} \textbf{89} 235101 (2014).

\bibitem{Sipahigil2014}
Sipahigil, A. \textit{et al.}
\newblock Indistinguishable Photons from Separated Silicon-Vacancy Centers in Diamond.
\newblock \textit{Phys. Rev. Lett.} \textbf{113} 113602 (2014).

\bibitem{juan2015}
Juan, M.L., Molina-Terriza, G., Volz, T., \& Romero-Isart, O.
\newblock Near-field Levitated Quantum Optomechanics with Nanodiamonds: Strong Single-Photon Coupling at Room Temperature
\newblock arxiv: 1505.03363 (2015).



\end{thebibliography}

\begin{thebibliography}{99}

\bibitem{SMBreunig2000}
M.~M.~Breunig, H.-P.~Kriegel, R.~T.~Ng, and J.~Sander,
\newblock LOF: identifying density-based local outliers,
\newblock \textit{SIGMOD Rec.} \textbf{29} 93-104 (2000).


\bibitem{SMGrynberg2010}
G.~Grynberg, A.~Aspect, and C.~Fabre,
\newblock \textit{Introduction to Quantum Optics From the Semi-classical Approach to Quantized Light}
\newblock Cambridge University Press, Cambridge (UK).


\bibitem{SMInam2011}
F. A. Inam, T. Gaebel, C. Bradac, L. Stewart, M. J. Withford, J. M. Dawes, J. R. Rabeau, and M. J. Steel, 
\newblock Modification of spontaneous emission from nanodiamond colour centres on a structured surface
\newblock \textit{New J. Phys.} \textbf{13}, 073012 (2011).

\bibitem{SMSiyushev2009}
P. Siyushev, V. Jacques, I. Aharonovich, F. Kaiser, T. M\"uller, L. Lombez, M. Atat\"ure, S. Castelletto, S. Prawer,
F. Jelezko, and J Wrachtrup,
\newblock Low-temperature optical characterization of a near-infrared single-photon emitter in nanodiamonds
\newblock \textit{New J. Phys.} \textbf{11}, 113029 (2009).

\bibitem{SMFaraon2011}
A.~Faraon, P.~E.~Barclay, C.~Santori, K.-M.~C.~Fu, and R.~G.~Beausoleil,
\newblock Resonant enhancement of the zero-phonon emission from a colour centre in a diamond cavity
\newblock \textit{Nat. Phot.} \textbf{5}, 301 (2011).

\bibitem{SMAlbrecht2013}
R.~Albrecht, A.~Bommer, C.~Deutsch, J.~Reichel, and C.~Becher,
\newblock Coupling of a Single Nitrogen-Vacancy Center in Diamond to a Fiber-Based Microcavity
\newblock \textit{Phys. Rev. Lett.} \textbf{110}, 243602 (2013).

\bibitem{SMHuxter2013}
V.~M.~Huxter, T.~A.~A.~Oliver, D.~Budker, and G.~R.~Fleming,
\newblock Vibrational and electronic dynamics of nitrogenÐvacancy centres in diamond revealed by two-dimensional ultrafast spectroscopy,
\newblock \textit{Nat. Phys.} \textbf{9} 744-749 (2013).


\bibitem{SMFu2009}
K.-M.~C.~Fu, \textit{et al.},
\newblock Observation of the Dynamic Jahn-Teller Effect in the Excited States of Nitrogen-Vacancy Centers in Diamond,
\newblock \textit{Pys. Rev. Lett.} \textbf{103} 265404 (2009).

\bibitem{SMHarada1995}
Y.~Harada, and T.~Asakura,
\newblock Radiation forces on a dielectric sphere in the Rayleigh scattering regime.
\newblock \textit{Optics Communications} \textbf{124} 529-541 (1995).

\bibitem{SMGieseler2013}
J. Gieseler, L. Novotny, and R. Quidant,
\newblock Thermal nonlinearities in a nanomechanical oscillator,
\newblock \textit{Nat. Phys.} \textbf{9} 806-810 (2013).



\bibitem{SMGross1982}
M.~Gross and S.~Haroche,
\newblock Superradiance: an essay on the theory of collective spontaneous emission
\newblock \textit{Physics Reports} \textbf{93} 301-396, (1982).


\end{thebibliography}
\end{document}